# Growth and characterization of heteroepitaxial La-substituted BaSnO$_3$ films on SrTiO$_3$ (001) and SmScO$_3$ (110) substrates.

P.V. Wadekar[1], J. Alaria[2], M. O'Sullivan[1], N.L.O. Flack[1], T.D. Manning[1], L.J. Phillips[2], K. Durose[2], O. Lozano[3], S. Lucas[3], J. B. Claridge[1], M.J. Rosseinsky[1].

[1] *Department of Chemistry, University of Liverpool, Crown Street, Liverpool L69 7ZE, UK.*

[2] *Stephenson Institute for Renewable Energy & Department of Physics, University of Liverpool, Crown Street, Liverpool L69 7ZD, UK.*

[3] *Research Centre for the Physics of Matter and Radiation, Namur Research Institute for Life Sciences, University of Namur, Namur 500, Belgium.*

Heteroepitaxial growth of BaSnO$_3$ (BSO) and Ba$_{1-x}$La$_x$SnO$_3$ (x = 7%) (LBSO) thin films on different perovskite single crystal (SrTiO$_3$ (001) and SmScO$_3$ (110)) substrates has been achieved by Pulsed Laser Deposition (PLD) under optimized deposition conditions. X-ray diffraction measurements indicate that the films on either of these substrates are relaxed due to the large mismatch and present a high degree of crystallinity with narrow rocking curves and smooth surface morphology while analytical quantification by proton induced x-ray emission (PIXE) confirms the stoichiometric La transfer from a polyphasic target, producing films with La contents above the bulk solubility limit. The films show degenerate semiconducting behavior on both substrates, with the observed room temperature resistivities, Hall mobilities and carrier concentrations of 4.4 mΩcm, 10.11 cm$^2$V$^{-1}$s$^{-1}$, and 1.38 × 10$^{20}$ cm$^{-3}$ on SmScO$_3$ and 7.8 mΩcm, 5.8 cm$^2$V$^{-1}$s$^{-1}$, and 1.36 × 10$^{20}$ cm$^{-3}$ on SrTiO$_3$ ruling out any extrinsic contribution from the substrate. The superior electrical properties observed on the SmScO$_3$ substrate are attributed to reduction in dislocation density from the lower lattice mismatch.

Tin doped indium oxide (ITO, $In_{2.95}Sn_{0.05}O_3$) has been extensively studied as a transparent conducting oxide (TCO) because of the unusual combination of high transparency in the visible range coupled with metal-like conductivity and high mobility. These properties make ITO important for optoelectronic devices such as flat panel displays, light emitting diodes and solar cells.[1] The high dispersion of the bottom of the s-type conduction band and the resulting low effective mass along with the lowered optical absorption due to hybridization between Sn and In *s* states produce the enhanced electrical and optical properties of ITO.[2] However, the scarcity of indium had led to the search for other doped oxide semiconductors such as ZnO,[3] $TiO_2$[4] and $SnO_2$[5,6] that can be used as alternatives to ITO with resistivity approaching the application threshold of $10^{-1}$ mΩcm[7] but device patterning and environmental stability issues are unresolved.[8,9] The rapid development of perovskite-based photovoltaic cells calls for compatible transparent electrodes.[10] Titanate-based perovskites have been heavily studied as potential TCO,[11-13] and recently, $BaSnO_3$ (BSO) has been proposed as a high performance perovskite TCO because its conduction band is composed of Sn 5s orbitals and La doping will create highly mobile carriers.[14] Initial work on thin films of doped stannates indicated that while oriented growth is possible, the resistivities are one order of magnitude higher[15-19] than found for ITO. Subsequent work on lanthanum doped barium stannate (LBSO) epitaxial thin films on $SrTiO_3$ (001)[20,21] substrates challenged this picture with room temperature mobility (resistivity) of ~ 70 $cm^2V^{-1}s^{-1}$, (0.17 mΩcm), which are comparable with conventional ITO thin films, making LBSO an alternative to ITO. The nominal La content of the films (x = 0.07) is beyond the bulk solubility limit, so determination of electrical properties for compositionally defined films is a priority. Such high mobility compared to the previous results is attributed to a superior crystalline quality of these thin films.[21] Although the apparent increase in mobility in the thin films is ascribed to the improved crystallinity, several other factors could be the origin of such results. Recent theoretical results highlight the possibility of tuning the bandgap using elastic strain,[22] and the interpretation of the electrical properties of films needs to take into account the role of any free carriers generated in the substrates. Oxygen vacancies generated during growth at high temperatures and low pressures in STO can produce conductivity.[23,24] All these potential problems make it difficult to robustly separate the films electrical properties from those of the substrate. Moreover it was suggested that the observed mobility in thin films could be improved by reducing the lattice mismatch between LBSO and the substrate.[20] In order to investigate these potential effects we have grown high quality BSO and LBSO films on $SrTiO_3$ (001) (STO) [lattice mismatch of +5.28%] and $SmScO_3$ (110) (SSO) substrates [+3.08% lattice mismatch ]. The scandate substrate possesses not only the largest lattice parameter of commercially available perovskite substrate to provides a lower mismatch for epitaxial growth but is also redox-resistant to introduction of carriers during processing, allowing for the investigation of possible strain effects and separation of the intrinsic electrical properties of the films.

Bulk targets of $BaSnO_3$ (BSO) and $Ba_{0.93}La_{0.07}SnO_3$ (LBSO) were prepared using $BaCO_3$, $La_2O_3$, and $SnO_2$ as the starting materials. Stoichiometric mixtures were ground, cold pressed and sintered at 1250 °C for 24 hours in alumina crucibles followed by grinding, after which the powders were cold isostatically pressed and fired at 1450 °C for 24 hours. XRD measurements showed that cubic perovskite phase with space group $Pm\bar{3}m$ was formed in the undoped BSO targets. However in case of LBSO, since the solid solubility of lanthanum is 3%[25], a polyphasic target with both perovskite $Ba_{1-x}La_xSnO_3$ and La-rich pyrochlore phase $La_2Sn_2O_7$ was formed with weight fractions of 95.7(3) and 4.3(2)% respectively. Epitaxial thin films of BSO and LBSO were grown on STO and SSO substrates using pulsed laser deposition with a 248 nm KrF laser. The substrates, STO (001) and SSO (110) were ultrasonically cleaned with acetone, ethanol and DI water and mounted on an Inconel sample holder with Ag paste. The vacuum chamber was pumped to a base pressure of $5 \times 10^{-7}$ Torr prior to deposition. Prior to deposition 100 mTorr of $O_2$ gas was introduced in the chamber and samples were annealed for 30 minutes. A systematic variation of heater temperatures from 700 $^o$C to 850 $^o$C and pressures from 100 mTorr to 2 mTorr was performed to find the optimal growth conditions. For all the depositions the laser energy was maintained at 180 mJ and frequency of 5 Hz. The numbers of pulses were adjusted to grow approximately 40 nm thick films. After growth, the samples were annealed for 10 minutes in 100 mTorr of $O_2$ gas and then cooled down to room temperature in the same oxygen pressure. Optimum growth was achieved for oxygen partial pressures of 2 mTorr and a heater temperature of 850 $^o$C. θ/2θ scans were measured in a two circle Panalytical X'Pert PRO diffractometer (Co $K_{α1}$), while rocking curves (RC), x-ray reflectivity (XRR) and reciprocal space maps (RSM's) were measured using a four circle Panalytical X'Pert PRO diffractometer (Cu $K_{α1}$). Surface morphology was studied using an Agilent 5600LS atomic force microscope (AFM). The temperature dependent electrical properties (resistivity, carrier concentration) were measured using van der Pauw geometry in a commercial Semimetrics 4C system using gold contacts. Rutherford back scattering (RBS) and proton induced x-ray emission (PIXE) measurements were performed in a Tandetron accelerator from High Voltage on thicker LBSO films grown under the same conditions as the film presented in this study in order to increase the volume probed. For RBS, a passive implanted planar silicon (PIPS) detector (Canberra) was used, while for PIXE, a low energy germanium (LEGe) detector (Canberra) was used. Protons of 2.5 MeV were used to bombard the samples positioned at 45°. Lanthanum quantification was performed in the L-shell energy window of 5.33-5.78 keV as the emission intensities are stronger.

PIXE spectra were analyzed by the software GUPIXWIN, while RBS where analyzed by SimNRA.[26]

Figure 1 (a) and (c) show the wide range XRD θ-2θ scans for the BSO/STO, LBSO/STO, BSO/SSO, LBSO/SSO film respectively. On either of these substrates, the (00$l$) peaks from the film are observed, indicating oriented growth along the c-axis. Diffraction peaks from secondary phases were absent. Similar diffraction peaks were also seen in case of the thicker LBSO/STO film. The bulk lattice constant of $BaSnO_3$ is 4.115(1) Å, while the measured out-of-plane lattice constant of BSO films on SSO and STO respectively are 4.117(2) Å and 4.122(1) Å, indicating that the undoped films are not completely relaxed. In case of single phase bulk lanthanum doped $BaSnO_3$, the lattice constant increases by 0.072% to 4.118(4) Å for $Ba_{0.97}La_{0.03}SnO_3$ (x = 3%).[25] A similar trend is also observed in the La-doped films where the lattice constant increases to 4.122(1) Å and 4.127(1) Å on SSO and STO respectively, an expansion of 0.12% compared to the undoped films. Assuming that the Vegard's law observed in the bulk is the same in the films, this lattice expansion would correspond to a doping level of x = 5% in the films. The crystalline quality is assessed by measuring the rocking curves of the (002) reflection for the doped samples. The top left inserts in figures 1(a) and 1(b) show the fitted RC. The measured full width half maximum (FWHM) for LBSO/STO and LBSO/SSO are 0.093° and 0.090° respectively which are comparable to the films of similar nominal compositions[20] making it possible to compare the electrical properties. The pendellosung fringes and size broadening (middle inserts in figure 1(a) and (c)) of the (002) Bragg peak of the films indicate that coherent growth with good crystallinity and smooth surfaces is sustained in the out-of-plane direction. Analysis of the Pendellosung fringes gives thicknesses of 37.1 nm on STO and 34.7 nm on SSO substrates. AFM measurements showed that films on either substrate were smooth with a root mean square roughness of less than 0.6 nm and are in agreement with the XRR model.[27] Asymmetrical RSM's were measured to ascertain the strain state in these films (Figure 1(b) and (d) for STO and SSO respectively). LBSO films on either of the substrates are not fully strained to the substrate due to the large mismatch, excluding the possibility of strain affecting the electrical properties. The more pronounced broadening in $Q_x$ for the film grown on STO compared to the film grown on SSO indicates the presence of a larger number of dislocations in the film. Although stoichiometric transfer from the target is often assumed to be achieved using PLD, as done in previous work on LBSO films, the analytical quantification of dopants in the actual films is crucial information in order to control and understand the electrical properties. RBS was used to determine the film thickness, while PIXE was used for chemical quantification, with the results plotted in figure 2(a) and 2(b) respectively. The film thickness was determined to be about 320 nm by simulating the RBS spectrum. PIXE data was analyzed using a variable width digital top hat filter function to suppress the background components and fitted using a least squares procedure. The atomic concentration of La was calculated as [Ba]$_{at.\ conc.}$ * La/[Ba]$_{mass}$ * ([Ba]$_{at.\ weight}$/ La$_{at.\ weight}$). The La content was 8±2% indicating that the targeted La doping level is achieved in the films, consistent with stabilization as films of compositions inaccessible in conventional bulk synthesis. If each La$^{3+}$ dopant produces one free electron, the expected carrier concentration in the Ba$_{0.93}$La$_{0.07}$SnO$_3$ film would be $9.8 \times 10^{20}$ cm$^{-3}$.

The electrical transport parameters (resistivity, carrier concentration and mobility) of the doped films are shown in Figure 3(a) and 3(b) for LBSO/SSO and LBSO/STO respectively. Hall effect measurements showed that the majority carriers in the films are electrons and have a room temperature effective carrier concentration of $1.38 \times 10^{20}$ cm$^{-3}$ for LBSO/SSO and $1.36 \times 10^{20}$ cm$^{-3}$ for LBSO/STO, indicating that the number of free carriers donated by La is the same irrespective of the substrates on which the films are grown. These values are lower than the carrier concentration expected assuming 1 electron/ La dopant and give a dopant activation rate of 14% which is much lower than the one reported previously.[20] The analytically determined carrier concentration is beyond the bulk solid solution limit and may drive enhanced carrier trapping associated with dopant clustering, producing the observed lower effective carrier concentrations.[28] The temperature dependence is consistent with a degenerate semiconductor with metallic-like behavior and follows the same trend in both cases, with carrier concentration decreasing as temperature is reduced as ionized dopant start to freeze-out. The films are more conducting on SSO than on STO (4.4 mΩcm and 7.8 mΩcm at room temperature respectively). Although similar FWHM are measured for the two films, dislocations that are formed due to lattice mismatch between film and substrate can reduce the mobility because they act as double Schottky barriers.[29] Although the carrier concentration is similar on the two substrates, a slight improvement in the lattice mismatch reduces the scattering by dislocations in the lower mismatched films thereby increasing the mobility, as seen for the films on STO (5.8 $cm^2 V^{-1}s^{-1}$) versus those on SSO (10.11 $cm^2 V^{-1}s^{-1}$). The mobilities increase as the temperature is decreased and saturate near 40 K suggesting that phonon scattering is the main process in this temperature regime. The mobilities obtained are consistent with the values expected for the total amount of impurity introduced in the lattice.[30]

In summary, we have successfully grown heteroepitaxial thin films of BSO and 7% La-doped BSO on STO and SSO substrates by PLD. Analytical chemical quantification by PIXE measurements indicates that the lanthanum doping concentration is 8±2% demonstrating that PLD gives higher La concentrations than possible in bulk ceramic synthesis. The influence of the substrate, specifically the lattice mismatch has been studied and we have shown that increased mobility can be achieved by lowering the misfit. The dopant activation rate in our film was found to be lower than the reported values and could be attributed to the samples being over-doped. The resulting compositionally-determined materials have electrical properties that are consistent across substrates with different mismatch and redox characteristics. The conductivities reported in this study, are close to alternative indium-free transparent conducting oxide thin films on both substrates and show that perovskite-compatible transparent electrodes can be developed.

This work is funded by the European Research Council (ERC Grant agreement 227987 RLUCIM).

[30]H. J. Kim, U. Kim, H. M. Kim, T. H. Kim, H. S. Mun, B. G. Jeon, K. T. Hong, W. J. Lee, C. Ju, K. H. Kim, and K. Char, Appl. Phys. Express **5**, 061102 (2012).

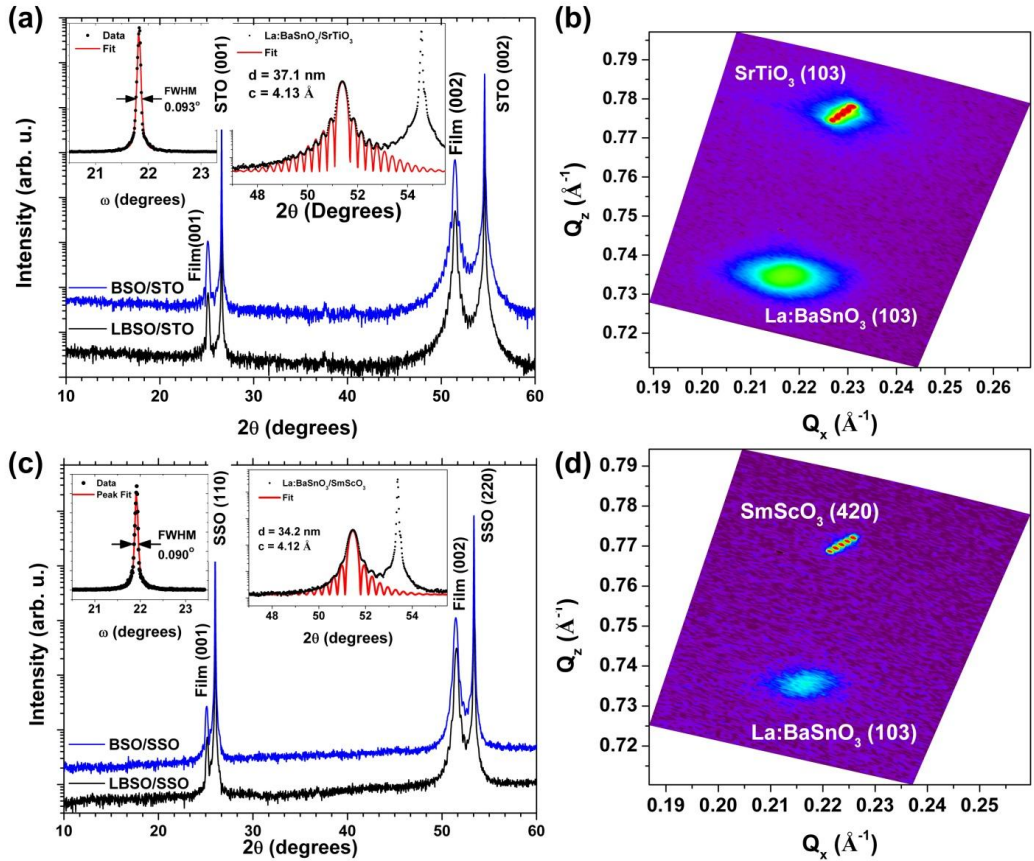

Figure 1(a) and (c):- θ-2θ scans for the BSO and LBSO films on STO and SSO substrates. Left inserts in (a) and (c) shows the rocking curve data with fitting of the full width half maximum, while the middle insert shows the pendellosung fringes with the associated fitting to derive the thickness. (b) Reciprocal space maps around STO (103) reflection (d) Reciprocal space maps around SSO (420) reflection.

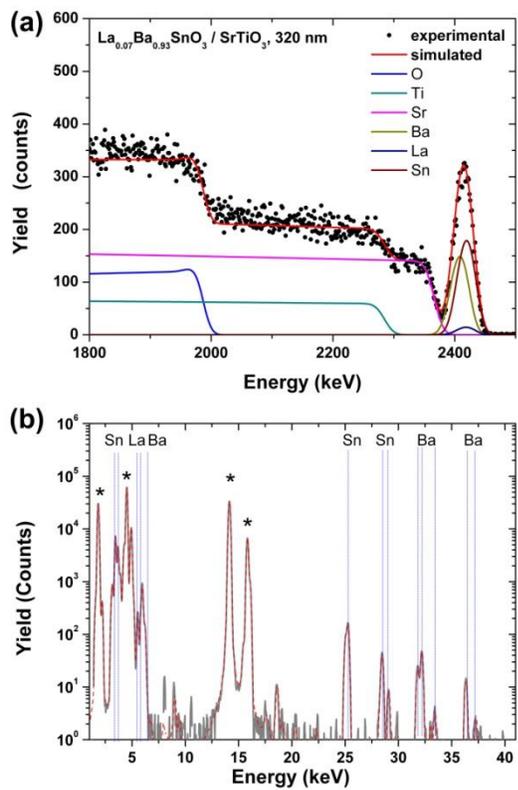

Figure 2 (a): Rutherford back scattering measurements for the thicker films along with the data fittings shown by red line. (b) PIXE spectra of the $La_{0.07}Ba_{0.93}SnO_3$ films on $SrTiO_3$ substrates. Dotted lines indicate emission from thin film elements. The asterisks (*) indicate emission from the substrate elements.

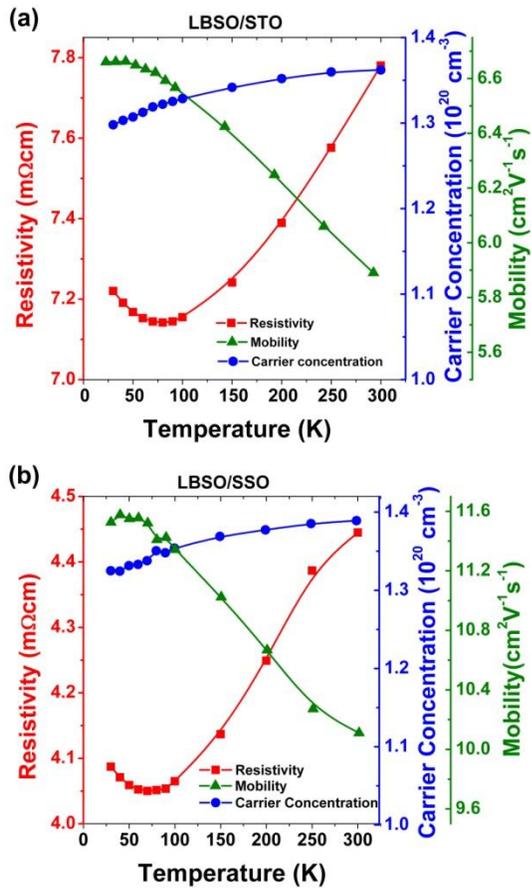

Figure 3(a) and (b):- Temperature dependent resistivity, mobility and carrier concentration measurements for LBSO films on STO and SSO substrates respectively.

Growth and characterization of heteroepitaxial La-substituted $BaSnO_3$ films on $SrTiO_3$ (001) and $SmScO_3$ (110) substrates.

P.V. Wadekar[1], J. Alaria[2], M. O'Sullivan[1], N.L.O. Flack[1], T.D. Manning[1], L.J. Phillips[2], K. Durose[2], O.Lozano[3], S.Lucas[3], J. B. Claridge[1], M.J.Rosseinsky[1].

[1] *Department of Chemistry, University of Liverpool, Crown Street, Liverpool L69 7ZE, UK.*

[2] *Stephenson Institute for Renewable Energy & Department of Physics, University of Liverpool, Crown Street, Liverpool L69 7ZD, UK.*

[3] *Research Centre for the Physics of Matter and Radiation, Namur Research Institute for Life Sciences, University of Namur, Namur 500, Belgium.*

# Supplementary information

In figure S1 (a) and (b), the x-ray reflectivity measurements along with a calculated model are plotted for the films on SrTiO$_3$ and SmScO$_3$ substrates respectively. The thicknesses (roughness) are 37 nm (0.45 nm) for SrTiO$_3$ and 34 nm (0.6 nm) for SmScO$_3$.

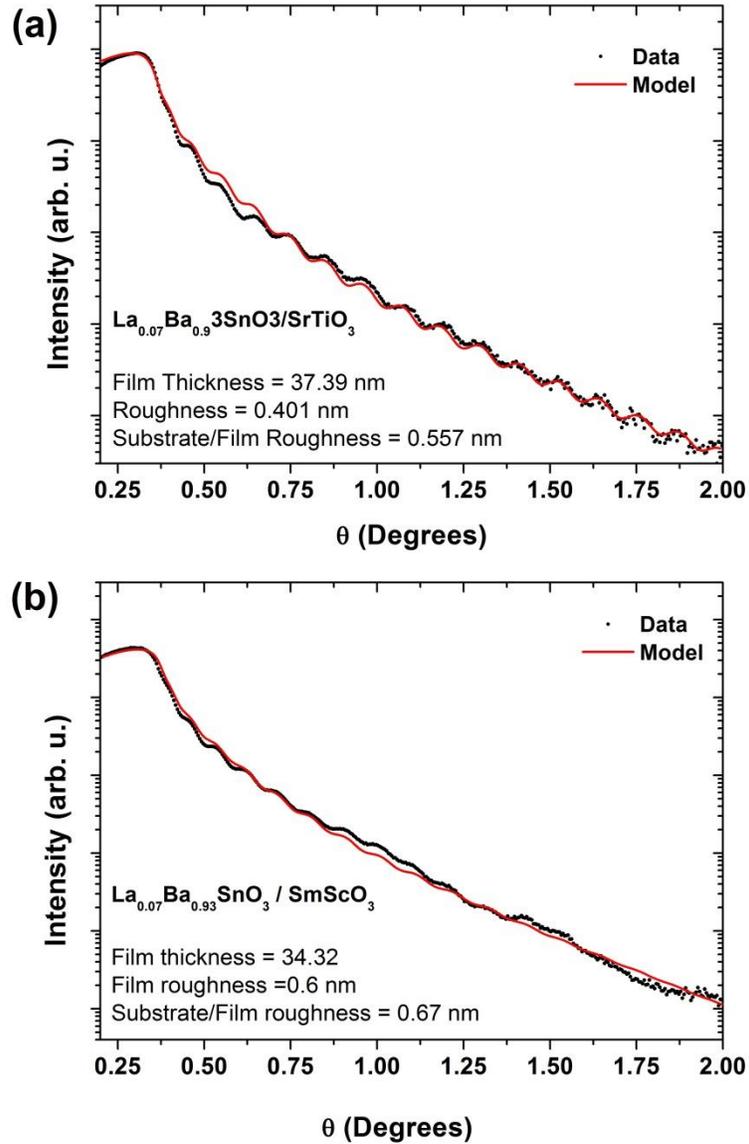

Figure S1: X-ray reflectivity measurements (black dots) and models (red line) for La$_{0.07}$Ba$_{0.93}$SnO$_3$ films on (a) SrTiO$_3$ and (b) SmScO$_3$ substrates

AFM morphology for the films is shown in figure S2 (a) and (b) for films on SrTiO$_3$ and SmScO$_3$ with the root mean square (RMS) roughness of 0.597 nm and 0.47 nm respectively

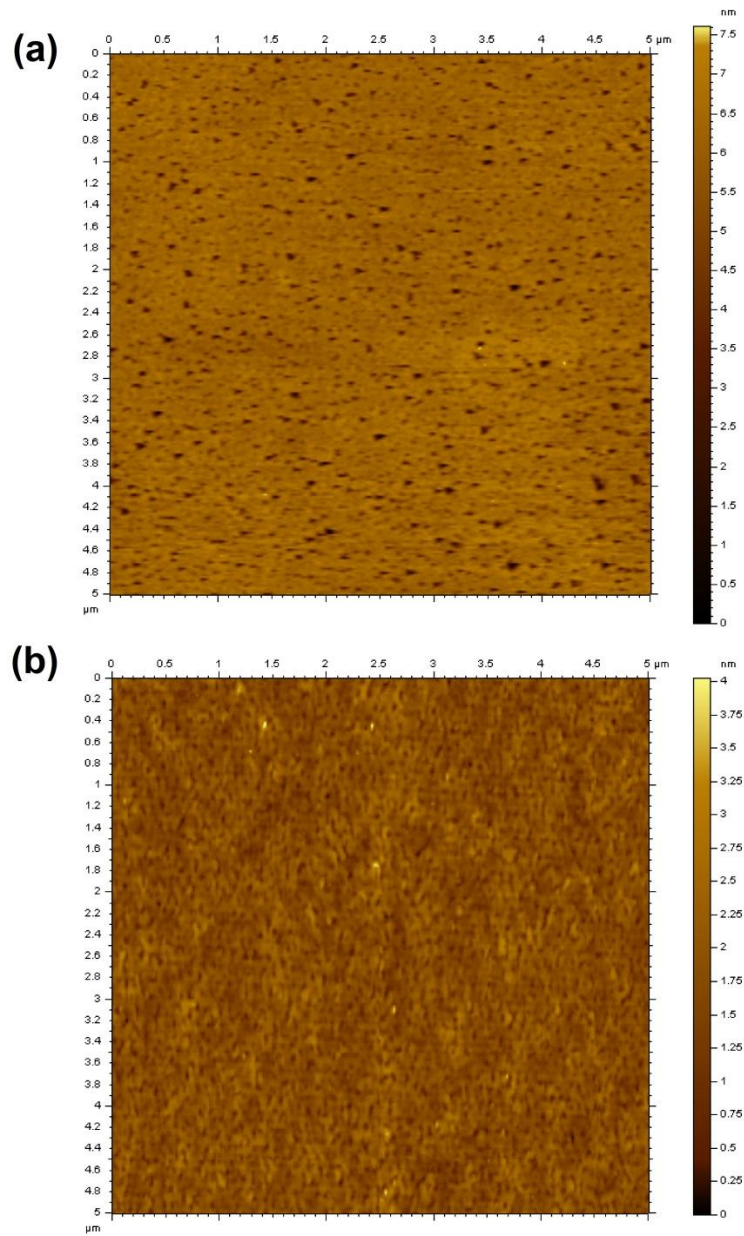

Figure S2: AFM surface morphology of the La$_{0.07}$Ba$_{0.93}$SnO$_3$ films grown on (a) SrTiO$_3$ and (b) SmScO$_3$ substrates